\documentclass[a4paper]{article}
\usepackage[utf8]{inputenc}
\usepackage[english]{babel}
\usepackage{authblk}
\usepackage{graphicx}
\usepackage{scrtime}
\usepackage{caption}
\usepackage[colorlinks]{hyperref}
\usepackage{etoolbox}
\usepackage[margin=3cm]{geometry}
\usepackage[finalnew]{trackchanges}

\usepackage{amsmath,amssymb}
\usepackage{algorithm,algpseudocode}

\RequirePackage[authoryear]{natbib}

\usepackage{parskip}

\begin{document}

\title{Zeitlin truncation of a Shallow Water Quasi-Geostrophic model for planetary flow}


\author[1]{Arnout Franken}
\author[2]{Martino Caliaro}
\author[1,2]{Paolo Cifani}
\author[1,3]{Bernard Geurts}

\affil[1]{Multiscale Modelling and Simulation, Department of Applied Mathematics, Faculty EEMCS, University of Twente, PO Box 217, 7500 AE Enschede, The Netherlands}
\affil[2]{Gran Sasso Science Institute, Viale F. Crispi 7, 67100 L'Aquila, Italy}
\affil[3]{Multiscale Physics, Center for Computational Energy Research, Department of Applied Physics, Eindhoven University of Technology, PO Box 513, 5600 MB Eindhoven, The Netherlands}

\date{}

\maketitle

\begin{abstract}
In this work, we consider a Shallow-Water Quasi Geostrophic equation on the sphere, as a model for global large-scale atmospheric dynamics. This equation, \add{previously} studied by \cite{Verkley2009a} and \cite{Schubert2009}, possesses a rich geometric structure, called Lie-Poisson, and admits an infinite number of conserved quantities, called Casimirs. In this paper, we develop a Casimir preserving numerical method for long-time simulations of this equation. The method develops in two steps: firstly, we construct an N-dimensional Lie-Poisson system that converges to the continuous one in the limit $N \to \infty$; secondly, we integrate in time the finite-dimensional system using an isospectral time integrator, developed by \cite{Modin2020}. We demonstrate the efficacy of this computational method by simulating a flow on the entire sphere for different values of the Lamb parameter. We particularly focus on rotation-induced effects, such as the formation of jets. In agreement with shallow water models of the atmosphere, we observe the formation of robust latitudinal jets and a decrease in the zonal wind amplitude with latitude. Furthermore, spectra of the kinetic energy are computed as a point of reference for future studies.   
\end{abstract}

\section{\label{sec:Introduction}Introduction}
In the context of Geophysical Fluid Dynamics, \add{the rotating stratified Euler equations constitute} an accurate model governing the dynamics of oceanic and atmospheric motions. \add{Due to their complexity, simpler models are often used to describe features of geophysical flows. Important examples of these are the Primitive Equations (PE) and shallow water models. In particular, the use of the Rotating Shallow Water (RSW) equations on the sphere has played a pivotal role in advancing our understanding of large-scale oceanic and atmospheric flows. The RSW equations describe a single shallow layer of fluid on a rotating sphere under hydrostatic pressure. The equations can be derived from the PE under the assumption of constant density and columnar motion and the subsequent vertical averaging of pressure gradients. The relation between these models can be found in more detail in} \cite{Zeitlin2018} \add{and} \cite{Luesink2021}.\\

From the one-layer global shallow-water model of the atmosphere, a simplified equation has been independently derived by \cite{Verkley2009a} and \cite{Schubert2009}, based on previous works of \cite{Charney1962} and \cite{Kuo1959}. This equation, which we will refer to as the Balanced Shallow Water (BSW) equation, is derived from RSW using a non-divergence assumption for the horizontal velocity field, together with a linear balance equation linking the stream function and the geopotential. This simplifying procedure leads to an equation for the potential vorticity that is identical in form to the rotating Euler equation for the relative vorticity, except for an additional $f^2$-dependent term, where $f$ is the Coriolis parameter. To validate the BSW model, Verkley considers the dynamics of linearized Rossby waves and shows that these waves accurately reproduce the westward-propagating waves of the second class of the original shallow-water equations, studied in \cite{Longuet1968}. Moreover, the associated 2D turbulence is studied in \cite{Verkley2009a} and in \cite{Schubert2009}, where the anisotropic Rhines barrier is derived.\\

The geometric structure is a \remove{remarkable }property that the BSW equation and the rotating Euler equations share. As we will show, both equations form a Lie-Poisson system and admit an infinite number of independent conserved quantities, called Casimirs. \add{These Casimirs can be expressed as integrated powers of potential vorticity, of which the integrated squared potential vorticity is called the enstrophy.}\\

\add{In a dynamical system, the presence of conserved quantities has profound physical consequences. If we consider the role of enstrophy in decaying 2D turbulence in the context of fluid dynamics, the well-known physical consequence is the inverse energy cascade. This phenomenon is opposite to the forward energy cascade that is observed in 3D turbulence, where enstrophy is not conserved (for the inviscid dynamics).}\\

\add{The presence of conserved Casimirs for 2D flows similarly has implications on the long-time behaviour of the fluid and for the formation of large-scale coherent structures} (\cite{Bouchet2012})\add{. These phenomena are sometimes explained through the equilibrium statistical theory. This theory typically makes use of some conserved quantities of the underlying (inviscid) dynamics to make predictions. For example,} \cite{Kraichnan1975}, \cite{Salmon1976} \add{and} \cite{Carnevale1987}\add{, derive equilibrium statistical states for the 2D Euler and the quasi-geostrophic equations using a Fourier-spectral truncation of the equations of motion. This spectral truncation preserves circulation, energy and enstrophy, but not the Casimirs of order higher than two. Therefore, it is of interest to investigate whether higher-order Casimirs are statistically relevant for geophysical equations.} \\

\add{This topic has been addressed in} \cite{Abramov2003} \add{for the barotropic two-dimensional flow with topography on a flat geometry (see also} \cite{Dubinkina2010}).\add{ In this work the authors use a Casimir preserving numerical method to integrate the equations of motion. In particular, they prove the statistical relevance of the third order Casimir $\mathcal{C}_3$ by showing an increasing discrepancy with the energy-enstrophy statistical theory (which predicts collinearity between the time-averaged streamfunction and potential vorticity) as a function of $\mathcal{C}_3$. We remark that, in order to study numerically the statistical relevance of high-order Casimirs in geophysical flows, it is necessary to have a numerical method that preserves the Casimirs in the discrete model.} \\

\add{Motivated by these considerations, the present work introduces a Casimir preserving numerical method for the BSW model on the sphere}, based on Zeitlin discretization \cite{Zeitlin1991, Zeitlin2004}, and on an isospectral time-integrator developed by \cite{Modin2020}. This method was recently applied successfully to underpin Kraichnan’s double scaling hypothesis for 2D forced turbulence \cite{Cifani2022b} and is also adopted here. \add{In this paper, we show that the combination of Zeitlin discretization and isospectral time-integration for the BSW flow on a sphere captures two important physical phenomena: the first is the double cascade in the kinetic energy spectrum as is typical for 2D turbulence; the second is the emergence of zonal jets across the sphere. In particular, the intensity of these zonal jets is not uniform across latitudes. We observe that the strong equatorial jets are attenuated towards the poles, and we will show that the intensity of this attenuation is regulated by a single parameter in this model, called Lamb's parameter. We remark that the attenuation of zonal jets, which is seen in physical systems such as the atmospheres of giant gaseous planets, has been observed in the simulations of RSW equations, but not in simulations of the 2D Navier-Stokes equations (see the work of} \cite{Scott2007}\add{ for the simulations of RSW where the attenuation of zonal jets is regulated by changing the Rossby deformation radius, similar to our case). For this reason, the BSW model may be a valuable tool for studying the formation and attenuation of zonal jets.} \\

The paper is organized as follows. In Section~\ref{sec:Equations} the mathematical model for BSW flow on a sphere is presented. Section~\ref{sec:Methods} \add{introduces} the Zeitlin truncation and the Lie-Poisson time-integration method. \add{At the end of this section we simulate an inviscid unforced BSW flow to demonstrate the numerical conservation of energy and the Casimirs.} Simulations of forced BSW flow in the presence of dissipation are performed and discussed in Section~\ref{sec:Experiment} and concluding remarks are collected in Section~\ref{sec:Conclusions}. 

\section{\label{sec:Equations}BSW equations on the sphere}

The one-layer rotating shallow water equation on the sphere is a model that, under the assumptions of constant density and columnar motion, describes the motion of a thin layer of fluid on the surface of a rotating sphere. The fluid is described by a two-dimensional horizontal (or tangential) velocity field $\textbf{u}(\phi,\theta)=(u,v)$ and a local height $h$\add{, where $\phi$ and $\theta$ are the latitude and longitude respectively}. This system admits a materially conserved quantity called potential vorticity: 
\begin{equation}
    P:=\frac{\omega+f}{h}
    \label{eq:pot_vort}
\end{equation}
where $\omega := \hat{\mathbf{r}} \  \cdot \ (\nabla \times \mathbf{u})$ is the radial component of the relative vorticity and $f = 2\Omega \sin \phi$ is the Coriolis parameter. Moreover, $\Omega$ denotes the angular velocity of the planet\remove{, and $\phi$ is the latitude}. \\ 

Starting from the one-layer RSW equations on the sphere, \cite{Verkley2009a} and \cite{Schubert2009} derived the simplified BSW model, by adding three assumptions as follows:\\

\textbf{Assumption 1}: We denote the local height of the fluid column by
\begin{equation}
    h(\mathbf{x},t) = H+\eta(\mathbf{x},t) + b(\mathbf{x})
\end{equation}
where $H$ is a uniform average height and $\eta$ is a variable surface elevation. The quantity $b$ denotes the topography and in the following will be assumed to be zero ($b(\mathbf{x}) = 0$). The first assumption consists of assuming that the variable surface elevation is much smaller than the average height, i.e.,
\begin{equation}
    \eta/H \ll 1
\end{equation}\\

\textbf{Assumption 2}: Motivated by this first assumption and by the continuity equation for RSW, we assume that the divergence of the horizontal velocity can be approximated as zero.
Using the Helmholtz decomposition, we write
the horizontal velocity in terms of a stream function $\psi$ such that 
\begin{equation}
    \mathbf{u} = \hat{\textbf{r}} \times \nabla \psi
    \label{assumption_2}
\end{equation}
\add{with $\nabla = \frac{1}{R}\frac{\partial (\cdot)}{\partial \phi}\hat{\phi} + \frac{1}{R\cos\phi}\frac{\partial (\cdot)}{\partial \theta}\hat{\theta}$ 
the horizontal gradient on the surface of the sphere with radius $R$.}\\

\textbf{Assumption 3}: In order to derive a closed dynamical system from the material conservation of the potential vorticity (\ref{eq:pot_vort}), we propose a type of balance that relates the velocity field $\textbf{u}$ to the local height $h$, or, equivalently to the \add{surface elevation $\eta$.} Among the linear balance relations, we choose the simplest one, which relates the stream function and the variable height as follows:
\begin{equation}
    f\psi = g \eta,
    \label{eq:balance}
\end{equation}\add{where $g$ is the gravitational acceleration.}
This relation is inspired by quasi-geostrophic systems, in which there exists an approximate balance between the Coriolis force and pressure forces. It was first considered by \cite{Daley1983}, as a simplification of the balance relation $\nabla \cdot(-f\nabla\psi + g\nabla \eta )=0$ introduced by \cite{Lorenz1960}.\\

Using Assumptions 1, 2 and 3 we can derive the BSW system from the RSW point of reference. Since we assume a small free surface elevation, an expansion up to first order in $\eta/H$ yields
\begin{equation}
    \frac{1}{h} = \frac{1}{H}\left(1-\frac{\eta}{H}\right)
\end{equation}
so that the potential vorticity becomes
\begin{equation}
    P = \frac{\omega+f}{H}\left(1 -\frac{\eta}{H}\right)
\end{equation}
In the regime of rapid rotation, the dominant contribution to the potential vorticity according to \cite{Verkley2009a} reads
\begin{equation}
    HP = \omega + f -\frac{\eta}{H}f =: q
    \label{eq:pot_vort_approx}
\end{equation}
defining the central dynamic variable $q$ to which we turn next. By using the linear balance relation (\ref{eq:balance}) we write the potential vorticity $q$ as:
\begin{equation}
    q = \omega + f -\frac{f^2}{gH}\psi 
\end{equation}
In this approximation, we can formulate the material conservation law for the potential vorticity (\ref{eq:pot_vort}) as 
\begin{equation}
    \frac{\partial q}{\partial t} + (\hat{\textbf{r}} \times \nabla \psi) \cdot \nabla q = 0,
    \label{eq:QG}
\end{equation}
where
\begin{equation}
    q = 2\Omega\sin\phi + \Delta \psi - \frac{4 \Omega^2 }{gH}\sin^2\phi \psi
    \label{eq:potential_vorticity}
\end{equation}
\add{where $\Delta:=\nabla\cdot\nabla$ is the Laplace-Beltrami operator on the surface of the sphere.}\\

Equations (\ref{eq:QG}) and (\ref{eq:potential_vorticity}) are the central formulation of the BSW model on the sphere. Anticipating the Zeitlin truncation~\cite{Zeitlin2004} in the next section, we non-dimensionalize the equations using the radius of the sphere $R$ as the length scale, and $\Omega^{-1}$ as the time scale. \add{Furthermore, we denote the convective term in} (\ref{eq:QG}) \add{as}
\begin{equation}
    \{\psi, q\} := -(\hat{\textbf{r}} \times \nabla \psi) \cdot \nabla q =  \frac{1}{\cos \phi}\left( \frac{\partial\psi}{\partial\phi}\frac{\partial q}{\partial\theta} - \frac{\partial \psi}{\partial\theta}\frac{\partial q}{\partial\phi}\right).
    \label{eq:PoissBracket}
\end{equation}
\add{The binary operation} $\{\cdot, \cdot\}$\add{ forms a Poisson bracket on the space of smooth functions on the sphere (i.e. the operation is bilinear, skew-symmetric and respects Jacobi and Leibniz identities} \cite{Arnold1992}). \add{Thus, equations} (\ref{eq:QG}) and (\ref{eq:potential_vorticity}) \add{can be reformulated as}
\begin{equation}
    \begin{cases}
        \dot{q} = \left\{ \psi,q \right\} \\
        \left(\Delta -\gamma \mu^2\right) \psi = q - 2\mu,
        \label{eq:Lie-Poisson}
    \end{cases}
\end{equation}
where $\mu = \sin{\phi}$
and $\gamma$ is Lamb's parameter, which expresses the ratio between the radius of the sphere, and the typical size of quasi-geostrophic vortices \cite{Vallis2019} given by the Rossby deformation length $R_d$:
\begin{equation}
    \gamma  = 4\frac{R^2}{R_d^2}, \quad  \mbox{where}\quad R_d = \frac{\sqrt{gH}}{\Omega}.
\end{equation}\\

\add{We now turn to the geometric properties of BSW equations in }(\ref{eq:Lie-Poisson}). \add{As anticipated, the BSW model possesses a rich geometric structure, which it inherits from the RSW model. Indeed, equations} (\ref{eq:Lie-Poisson}) \add{constitute an infinite-dimensional Lie-Poisson system} \cite{Marsden1983,Arnold1992} \add{on the space of smooth functions on the sphere $C^\infty(\mathbf{S}^2)$. In other words, the BSW equations have a Hamiltonian structure with an additional Lie algebra property.} \\

\add{The Hamiltonian structure is defined as follows. Firstly, the space $C^{\infty}(\mathbf{S}^2)$, i.e. the space of vorticity fields, forms the infinite dimensional phase space for} (\ref{eq:Lie-Poisson})\add{. Secondly, the Poisson bracket in} (\ref{eq:PoissBracket}) \add{defines a new infinite dimensional Poisson bracket $\{\cdot,\cdot\}_{LP}$ that acts on functionals defined on the phase space as follows:}
\begin{equation}
    \{\mathcal{F},\mathcal{G}\}_{LP}(q) := \int_{\mathbf{S}^2} q \left\{ \frac{\delta \mathcal{F}}{\delta q}(q), \frac{\delta \mathcal{G}}{\delta q}(q)\right\} \, dA,
\end{equation}
\add{for smooth functionals $\mathcal{F},\mathcal{G}: C^{\infty}(\mathbf{S}^2) \to \mathbf{R}$. Here $dA =-\cos\phi d\phi d\theta$ is the infinitesimal surface element, and $\delta(\cdot)/\delta q$ denotes the variational derivative. When, as in} \cite{Verkley2009a},  \add{we define the Hamiltonian of BSW as}
\begin{equation}
    \mathcal{H}(q) = -\frac{1}{2}\int_{\mathbf{S}^2} (q-2\mu)\psi\,dA,
\end{equation} 
\add{the potential vorticity equation} (\ref{eq:Lie-Poisson}) \add{can be written in equivalently as}
\begin{equation}
    \frac{d}{dt}\mathcal{F}(q) = \{ \mathcal{F}(q), \mathcal{H}(q) \}_{LP}, \ \ \ \forall\mathcal{F}:C^{\infty}(\mathbf{S}^2) \to \mathbf{R}.
    \label{eq:hamilton}
\end{equation} 
\add{In fact, in the case $\mathcal{F}(q) = q(\Vec{x})$, for $\Vec{x} \in \mathbf{S}^2$ a given point on the sphere, we recover} (\ref{eq:Lie-Poisson}).\\ 

\add{The additional Lie algebra property arises from the following observation. Since the phase space $C^{\infty}(\mathbf{S}^2)$ is a vector space, the couple $(C^{\infty}(\mathbf{S}^2), \{\cdot, \cdot\})$ forms an infinite dimensional Lie algebra} \cite{Olver1993}. \add{This additional property implies the existence of special functionals $\mathcal{C}(q):C^{\infty}(\mathbf{S}^2) \to \mathbf{R}$ with the property}
\begin{equation}
     \{ \mathcal{C}(q), \mathcal{F}(q) \}_{LP} = 0 \ \ \  \ \ \ \ \ \forall \mathcal{F}:C^{\infty}(\mathbf{S}^2) \to \mathbf{R}.
     \label{property_casimir}
\end{equation}
\add{These functionals are called Casimirs and, due to equation} (\ref{eq:hamilton}), \add{are constants of motion. The Casimir functions for the BSW model are an infinite family and are given, for any smooth function $\xi \in C^\infty(\mathbf{R})$, by}
\begin{equation}
    \mathcal{C}(q) = \int_{\mathbf{S}^2} \xi(q)\,dA
    \label{eq:casimir}
\end{equation}
\add{In practice, the functions $\xi(q)$ can be chosen to be monomials, so the corresponding Casimirs read}
\begin{equation}
    \mathcal{C}_k(q) = \int_{\mathbf{S}^2} q^{k}\,dA \ \ \ \text{for} \ \ k=1,2,...
\end{equation}
\add{The conservation of Casimirs for BSW can also be verified via direct computation:}
\begin{equation}
   \begin{split}
        \frac{d}{dt}\mathcal{C}(q) &= \frac{d}{dt} \int_{\mathbf{S}^2} \xi(q)\,dA = -\int_{\mathbf{S}^2} \xi'(q) \mathbf{u}\cdot \nabla q \,dA = -\int_{\mathbf{S}^2}  \mathbf{u}\cdot \nabla \xi(q) \,dA \\ & = \int_{\mathbf{S}^2}  (\nabla \cdot \mathbf{u}) \xi(q) \,dA = 0
   \end{split}
\end{equation}
\add{where the second equality uses equations} (\ref{assumption_2}) and (\ref{eq:QG}). \add{Notably, the conservation of the Casimirs is independent of the particular relation between $q$ and $\psi$. In other words, their conservation is independent of the choice of the Hamiltonian, as equation} (\ref{property_casimir}) \add{suggests. This reflects the underlying Lie-Poisson structure. In the next section, we will derive a special finite-dimensional truncation of equations} (\ref{eq:Lie-Poisson}) \add{that preserves this Lie-Posson structure and, in particular, that preserves all Casimirs of the discrete model.}\\

\add{We conclude by commenting on the geometric structure of BSW. As mentioned, this structure is inherited from the RSW model, which also exhibits a Lie Poisson structure and the conservation of infinitely many Casimirs (see, for example,} \cite{Salmon2004} and \cite{Dellar2005} \add{for a geometric formulation of RSW). This fact is not exceptional, as many other approximate dynamical equations exhibit geometric properties analogous to the exact parent cases} (see \cite{Salmon1988} \add{and references therein). For this reason, it would be interesting to investigate the possibility of deriving the BSW model utilizing the Hamiltonian formulation of the RSW equations, and this topic will be left for future research.}

\section{\label{sec:Methods}Numerical method for Zeitlin truncation}

In this section, we follow the approach by Zeitlin to construct a numerical scheme with which the geometric structure of the BSW equations can be preserved optimally -- this procedure is sometimes referred to as `quantization', contrasting the familiar `discretization'. Specifically, we show how to construct and integrate a finite-dimensional analogue of equations (\ref{eq:Lie-Poisson}) while preserving all the associated independent Casimirs. \add{The first part of} Subsection~\ref{subsec:finite_truncation} \add{is quite technical, and may be skipped upon first reading, continuing after the bullet points.}

\subsection{\label{subsec:finite_truncation}Finite truncation of the Poisson bracket}

We begin our construction of the Casimir-preserving numerical method by constructing a finite-dimensional analogue of (\ref{eq:Lie-Poisson}). This will be referred to as the Zeitlin truncation. Fundamental to the derivation, we notice that the couple $(C^{\infty}(\mathbf{S}^2), \{\cdot,\cdot\})$ forms an infinite dimensional Lie algebra and the existence of Casimirs is a direct result of this Lie-Poisson structure. Therefore, following Zeitlin, we look for a sequence of N-dimensional Lie algebras $(\mathbf{g}_N,[\cdot,\cdot]_N)$ that constitute an approximation, in a sense specified later, of $(C^{\infty}(\mathbf{S}^2), \{\cdot,\cdot\})$ in the limit $N \to \infty$. Here, $\mathbf{g}_N,$ denotes an N-dimensional vector space and $[\cdot,\cdot]_N$ refers to the bilinear form of the Lie algebra. The family of finite-dimensional Lie algebras  $(\mathbf{g}_N,[\cdot,\cdot]_N)$ is referred to as $L_{N}$-approximation\cite{Bordemann1991}.\\

In \cite{Bordemann1994}, it is shown that the $L_N$-approximation for $(C^{\infty}(\mathbf{S}^2), \{\cdot,\cdot\})$ can be obtained using the Lie algebra $(\mathfrak{gl}(N),[\cdot,\cdot])$, where $\mathfrak{gl}(N)$ is the set of complex $N \times N$ matrices and $[\cdot,\cdot]$ is the matrix commutator. Indeed, we can construct a surjective projection $\Pi_N: C^{\infty}(\mathbf{S}^2) \to \mathfrak{gl}(N)$ such that
\begin{itemize}
        \item $\Pi_N f - \Pi_N g \to 0$ as $N \to \infty$ implies $f = g$;
        \item $\Pi_N \{f,g\} = \frac{N^{3/2}}{\sqrt{16 \pi}}[\Pi_Nf,\Pi_Ng] + O(1/N^2)$
\end{itemize}
\add{This allows us to construct the finite-dimensional analogue of equations} (\ref{eq:Lie-Poisson}). Using the projection \add{$\Pi_N$}, we associate with every smooth function on the sphere $\mathbf{S}^2$ a \add{complex $N \times N$} matrix in \add{the Lie algebra} $\mathfrak{gl}(N)$. The action of the projection $\Pi_N$ is linear and explicitly known: to each spherical harmonics $Y_{l,m} \in C^{\infty}(\mathbf{S}^2)$ we associate a matrix $i\widehat{T}_{l,m} \in \mathfrak{gl}(N)$. Here $i$ is the imaginary unit and the explicit expression of the matrices $\widehat{T}_{l,m}$ is reported in Appendix A. Thus, using the spherical harmonics as a basis, we associate with any function given by
\begin{equation}
    q = \sum_{l=0}^{\infty}\sum_{m=-l}^l q_{l,m}Y_{l,m} \in C^{\infty}(\mathbf{S}^2)
\end{equation}
the matrix
\begin{equation}
    Q:= \Pi_N q = \sum_{l=0}^{N-1} \sum_{m=-l}^{l} i q_{l,m} \widehat{T}_{l,m}^N.
\end{equation}
Notice that for a real-valued function $q$, the spherical harmonics coefficients satisfy $q_{l,m} = (-1)^mq_{l,-m}$. This translates to the matrix condition $Q + Q^\dagger=0$, where $Q^\dagger$ is the conjugate transpose of $Q$. For this reason, we can restrict to the subalgebra of skew-Hermitian matrices $Q \in \mathfrak{u}(N) \subset \mathfrak{gl}(N)$. 
In terms of the discrete basis, our system of equations (\ref{eq:Lie-Poisson}) takes the following form:
\begin{equation}
\begin{cases}
    \dot{Q} = \left[P,Q\right]_N\\
    \Delta_N P - \gamma\Pi_N (\mu^2 \psi ) = Q - 2\Pi_N \mu,
    \end{cases}
    \label{eq:Lie-Poisson_quantized}
\end{equation}
where $Q$ is usually referred to as the potential vorticity matrix and $P$ as the stream matrix. \add{Here, have replaced the Poisson brackets $\{\cdot,\cdot\}$ in} (\ref{eq:Lie-Poisson}) \add{with the }scaled\add{ matrix} commutator $[\cdot,\cdot]_N=N^{3/2}/\sqrt{16\pi}[\cdot,\cdot]$ as given by the aforementioned projection of the Poisson bracket \cite{Modin2020}. \add{Moreover, we have replaced the Laplace-Beltrami operator $\Delta$ in} (\ref{eq:Lie-Poisson}) \add{with} the quantized Laplace-Beltrami operator $\Delta_N$, that satisfies $\Delta_N \widehat{T}_{l,m} = -l(l+1)\widehat{T}_{l,m}$, as further detailed in \ref{app:basis}.\\

\subsection{Computation of the stream function}

The scheme presented in equations (\ref{eq:Lie-Poisson_quantized}) involves computation of the stream matrix $P$. Determining the stream matrix involves quantizing \add{(i.e. projecting using $\Pi_N$)} the functions $\mu$ and $\mu^2\psi$. For the projection of the former, we use the expansion in terms of spherical harmonics:
\begin{equation}
    \mu = \sin \phi = 2 \sqrt{\frac{\pi}{3}} Y_1^0.
\end{equation}
The discrete representation is therefore given by the following projection:
\begin{equation}
    M := \Pi_N \mu = 2i \sqrt{\frac{\pi}{3}} \widehat{T}_{1,0},
    \label{eq:M}
\end{equation}
which can be calculated explicitly. The term $\mu^2\psi$ is more involved, as it requires the projection of a product of functions. In \ref{app:product_projection}, we discuss a method for approximating the projection of a product of two functions on the sphere, together with its accuracy. In practice, given two real functions $f$ and $g$ and their associated matrices in $\mathfrak{u}(N)$ given by $F=\Pi_N f$ and $G = \Pi_N g$, we use the following approximation:
\begin{equation}
    \Pi_N (fg) \approx -\frac{i}{2} \sqrt{\frac{N}{4\pi}}(FG + GF),
    \label{eq:approx}
\end{equation}
which associates multiplication with the anti-commutator on $\mathfrak{u}(N)$. In our specific case, using the expansion:
\begin{equation}
    \sin^2 \phi = \frac{2}{3}\sqrt{\pi}Y_0^0  + \frac{4}{3} \sqrt{\frac{\pi}{5}} Y_2^0,
\end{equation}
we can define
\begin{equation}
    S:= \Pi_N \mu^2 = \frac{2i}{3}\sqrt{\pi}\widehat{T}_{0,0}^N + \frac{4i}{3} \sqrt{\frac{\pi}{5}} \widehat{T}_{2,0}^N.
\end{equation}
which enables writing:
\begin{equation}
    \Pi_N (\mu^2 \psi) \approx - \frac{i}{2} \sqrt{\frac{N}{4\pi}}(SP+PS).
\end{equation}\\

The final step which allows us to solve the system for $P$ is the observation that the matrix harmonics $\widehat{T}_{l,0}^N$ only have non-zero entries on the diagonal (see appendix \ref{app:basis}). If we denote the diagonal of $S$ as the vector $s\in \mathbb{C}^N$ such that $s_i=S_{ii}$, we can write:
\begin{equation}
    \Pi_N (\mu^2\psi) \approx \Tilde{S} \circ P, \quad \mbox{where} \quad \Tilde{S}_{ij} = -\frac{i}{2}\sqrt{\frac{N}{4\pi}}(s_i+s_j)
    \label{eq:Stilde}
\end{equation}
where $\circ$ is the Hadamard, or entry-wise product.\\

Finally, we arrive at the discrete system by substituting the above results into equations (\ref{eq:Lie-Poisson_quantized}):
\begin{equation}
    \begin{cases}
        \Dot{Q} = \left[ P, Q \right]_N \\
        \Delta_N P -\gamma \Tilde{S}\circ P = Q - 2 M
    \end{cases}
    \label{eq:zeitlin}
\end{equation}
where the matrices $M$ and $\Tilde{S}$ are defined in equations (\ref{eq:M}) and (\ref{eq:Stilde}) respectively. The equation for $P$ can be solved efficiently using a clever splitting of the problem along diagonals of the matrix $P$. The implementation is based on the methodology used by \cite{Cifani2023a} and is further detailed in \ref{ap:tridiagonal}.\\

The time-evolution of $Q$ in equation (\ref{eq:zeitlin}) is isospectral since the spectrum of the matrix is preserved by the matrix commutator \cite{Viviani2020}. This implies that traces of powers of the quantized potential vorticity are conserved. In other words, the quantities
\begin{equation}
    C_k(W) = \mbox{Tr} (Q^k) \ \ \text{for} \ \ k=1,2, ..., N
    \label{eq:discr_casimir}
\end{equation}
are conserved by (\ref{eq:zeitlin}). This represents the discrete analogue of the conserved Casimirs \cite{Cifani2023a} in equation (\ref{eq:casimir}). \add{In this case the conservation law for the Casimirs follows directly from the cyclic property of the trace.} \\ 

Additionally, the system is Lie-Poisson with the Hamiltonian $H_N$ given by:
\begin{equation}
    H_N = \frac{1}{2} \mbox{Tr} \left(P(Q-2M)^\dagger \right)
\end{equation}
System (\ref{eq:zeitlin}) provides a finite-dimensional analogue of equations (\ref{eq:Lie-Poisson}), relating the potential vorticity matrix to the stream matrix. In the next section, we integrate (\ref{eq:zeitlin}) by means of an isospectral time integrator that preserves all the Casimirs \add{in equation} (\ref{eq:discr_casimir}).

\subsection{Isospectral time-integration}

In order to preserve the discrete Casimirs (\ref{eq:discr_casimir}), we use an isospectral Lie-Poisson time-integrator following \cite{Modin2020}. This retains the basic discrete Lie-Poisson dynamics, \add{and, moreover, provides near-conservation of the Hamiltonian as obtained through backward error analysis} \cite{Hairer2006}. The time integrator is based on the implicit midpoint rule and can be written as follows:
\begin{equation}
    \begin{cases}
    \Tilde{Q} = Q_n + \frac{h}{2}[P,\Tilde{Q}]_N + \frac{h^2}{4}P\Tilde{Q} P \\
    Q_{n+1} = \Tilde{Q} + \frac{h}{2}[\Tilde{P},\Tilde{Q}]_N - \frac{h^2}{4}\Tilde{P}\Tilde{Q}\Tilde{P}
    \end{cases}
    \label{iso-algor}
\end{equation}
The first equation is solved for $\Tilde{Q}$ using fixed point iteration that creates a sequence $\{\Tilde{Q}_k\}$, with $\Tilde{Q}_0=Q_n$, which converges to $\Tilde{Q}$ as $k \rightarrow \infty$. This contraction was found to yield rapid convergence and only a small number of iterations was found necessary to reach a set tolerance. In test calculations we may typically reach an $L_\infty$-norm of $\Tilde{Q}_{k+1}-\Tilde{Q}_{k}$ of $O(10^{-12})$ after only $3$ iterations. To complete a time-step, the sufficiently converged approximation $\Tilde{Q}_K$ of $\Tilde{Q}$ for suitable $K$ is used once again to evaluate the full right-hand side in the second equation in (\ref{iso-algor}). \\

\add{The computational effort associated with the integration scheme is dominated by the evaluation of the matrix commutator. As shown in} \ref{ap:tridiagonal} \add{, the stream matrix can be evaluated very efficiently from the vorticity matrix, requiring only $O(N^2)$ operations. Evaluating the commutator, however, involves dense matrix multiplications, requiring $O(N^3)$ operations. This shows a comparable computational complexity to that of typical pseudo-spectral methods since transformations involving spherical harmonics require at least $O(N^2\log^2(N))$ operations, though typical implementations display $O(N^3)$ scaling} \cite{Schaeffer2013}.\add{ Additionally, the current scheme allows for an efficient MPI parallelisation as shown in }\cite{Cifani2023a}. \add{ Similar to pseudo-spectral methods, the memory requirements for this scheme are $O(N^2)$, associated with the storage of several dense complex matrices or vectors of spherical harmonics coefficients. Hence, the complexity of the new method is comparable to popular methods in literature, but in addition, the new method retains the geometric structure of the equations.}

\subsection{Numerical performance}

\add{To test the performance of the numerical method, we set up a numerical experiment in which the flow develops freely from an initial condition. In the absence of any forcing or dissipation, the Hamiltonian energy and all Casimirs remain constant at their respective initial values. We solve the system in equations} (\ref{eq:zeitlin}) \add{on the unit sphere using the integration scheme in equations }(\ref{iso-algor})\add{. The angular velocity of the sphere is set to $\Omega=250$, corresponding to approximately 40 revolutions per second, and the Lamb parameter is set at $\gamma = 10^3$. The rotation axis is placed vertically. We simulate the system using $N=512$, which represents a truncation in the expansion of spherical harmonics up to degree $l=511$. The system is solved up to $t=500$, corresponding to around $20\,000$ revolutions of the unit sphere. The time step is set at $\Delta t = 4\cdot 10^{-4}$, corresponding to around 60 time steps per revolution of the sphere, or, 'day'. }\\

\begin{figure}
\centering
\includegraphics[width=.7\textwidth]{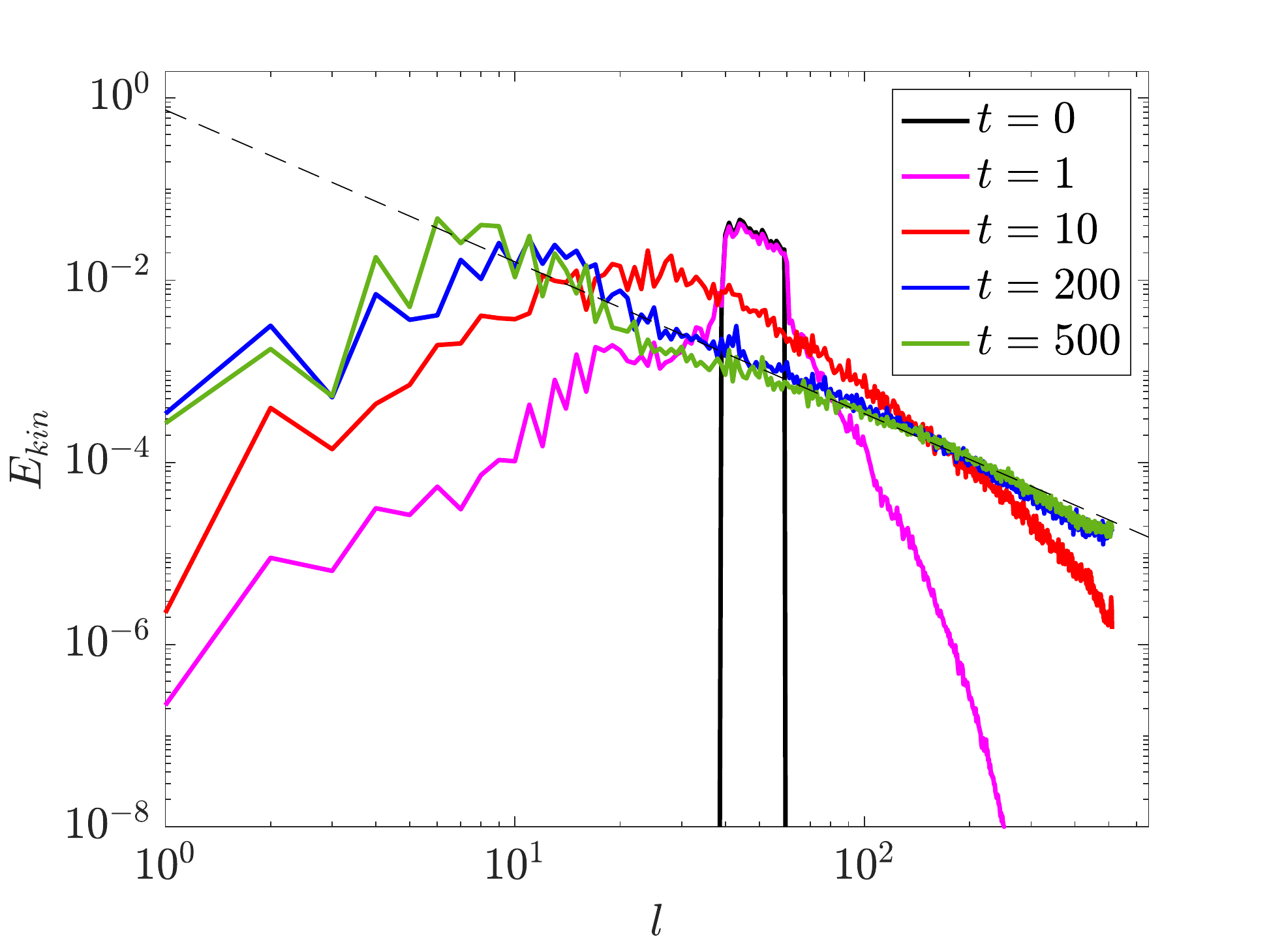}
\caption{\add{Kinetic energy spectrum of freely decaying turbulence on the sphere at several times. For reference, the slope -5/3 is shown as a dashed line.  }}
\label{fig:Spectrum_hyperbolic}
\end{figure}

Figure~\ref{fig:Spectrum_hyperbolic} \add{shows the evolution of the kinetic energy spectrum in time. As an initial condition, all modes between $l=40$ and $l=60$ are given a random amplitude around the reference value of $50/l(l+1)$. As the flow evolves in time, energy spreads to other modes through nonlinear interaction, with energy going to the larger scales as well as smaller scales. After this initial mixing phase, the spectrum reaches a statistically stationary state, as in} \cite{Modin2022}\add{. As a reference, the dashed line shows the -5/3 scaling for energy cascade in two-dimensional flows.} \\

\add{For any discretization of two-dimensional flows, the forward cascade of enstrophy is limited by the smallest length scales represented by the discrete system, leading in many cases to spectral blocking} \cite{Boyd2001}\add{. In this case, where the spatial resolution is determined by truncation of spherical harmonic modes on the sphere, spectral blocking occurs at the modes of the highest degree. At long simulation times, this does not seem to lead to instabilities of the numerical method and we find a similar trend as was observed for a Zeitlin truncation of the Euler equations by }\cite{Modin2022}\add{. They report that after an initial strong cascade of enstrophy to small scales, the system develops to statistically steady dynamics. Furthermore, they show that the resulting flow can be decomposed into large-scale dynamics over a noisy background, and they study the long-time behaviour of these large-scale structures (see also} \cite{Modin2020}). \add{This noisy background can be linked to the presence of the aforementioned spectral blocking effect.} \\

\begin{figure}
\centering
\includegraphics[width=\textwidth]{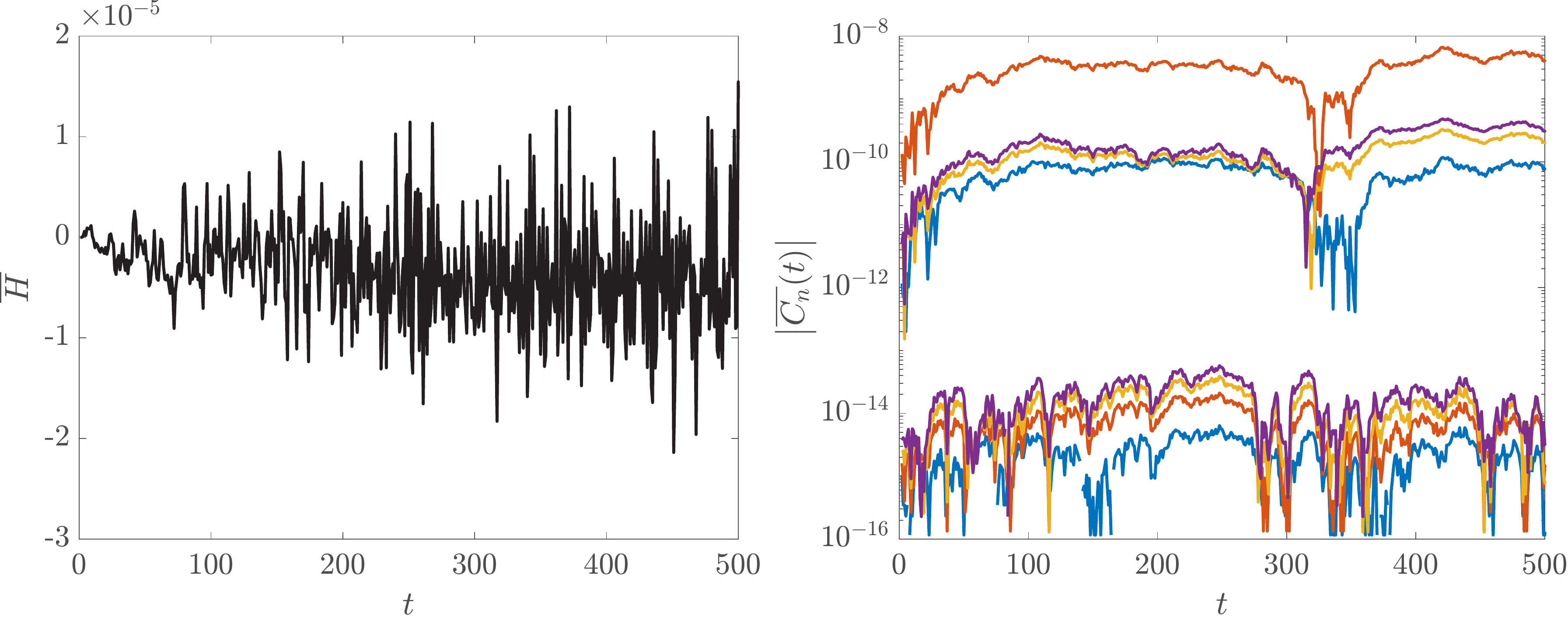}
\caption{\add{Time-evolution of the relative error $\overline{H}$ in the Hamiltonian as a function of time (left) and the error in the first 8 Casimirs of the discrete system, also normalized by their respective initial values (right). The Casimirs are grouped by the parity of their index. Casimirs with an odd index are clustered at the top of the figure, while Casimirs with an even index appear at the bottom of the figure.}   }
\label{fig:Conservation}
\end{figure}

\add{ In } Figure~\ref{fig:Conservation} \add{ the relative drift of the Hamiltonian and the first 8 Casimirs of the discrete system are shown. The left panel shows the evolution of the error Hamiltonian, normalized by its initial value, given as $\overline{H} = (H(t)-H(0))/H(0)$. As expected, the Hamiltonian is approximately conserved, i.e., it fluctuates close to the exact value with minimal linear drift. On the right panel, the relative error in the first 8 Casimirs are shown as defined in equation} (\ref{eq:discr_casimir}) \add{, where the relative error is given by $\overline{C}_n = (C_n(t)-C_n(0))/C_n(0)$. Although exact Casimir conservation is by the isospectrality of the time-integrator, it is in practice limited by the tolerance set for the iterative solver for equations} (\ref{iso-algor})\add{, which is set to $10^{-12}$ for this simulation. The figure shows that the relative error in the Casimirs stays below $10^{-10}$ even up to long times, with negligible drift, with Casimirs with an even index even showing conservation up to $10^{-14}$.}\\ 

\add{These results show that simulations of freely decaying turbulence of the BSW equation on a sphere with an energy and Casimir preserving method are indeed feasible and numerically stable. The effect of spectral blocking of the forward enstrophy cascade will be the subject of future research. In the next section, we will focus on forced turbulence in the presence of viscous dissipation.}

\section{\label{sec:Experiment}Simulation of zonal jet formation}

To illustrate the Zeitlin truncated BSW model with Casimir-preserving iso-spectral time-integration, we simulate a particular rotating sphere case. Results obtained using the BSW equation will be compared with findings based on the Navier-Stokes equations on a rotating sphere from \cite{Cifani2022b}. Therefore, we consider the BSW equation with additional forcing, dissipation and (Rayleigh) friction:
\begin{equation}
\begin{cases}
    \dot{q} = \left\{ \psi,q \right\} + \nu (\Delta q + 2 q) - \alpha q + F,  \\
    \left(\Delta - \gamma\mu^2\right) \psi = q - 2\mu,
\end{cases}
\label{eq:QGwithNS}
\end{equation}
where $\nu$ is the dimensionless viscosity, $F$ is the forcing and $\alpha q$ is the dimensionless Rayleigh friction to avoid an accumulation of energy at large scales due to the inverse energy cascade in 2D turbulence. Moreover, $2\nu q$ arises from the spherical geometry \cite{Lindborg2022}. Following \cite{Cifani2022b}, the forcing is time-dependent, derived from a Wiener process uncorrelated to the time scales of the flow. The forcing is localized in a narrow band in spherical harmonic space around degree $l_f=100$.\\

The viscous dissipation, damping and forcing are integrated using a second-order Crank-Nicolson scheme, while the isospectral integrator is used for the convective term. This ensures that the solution departs from the level sets of the Casimirs only due to the non-conservative terms in the equation and not by numerical inaccuracies in the discretization of convection. By denoting with $\Phi_{iso,h}$ the isospectral map for convection and by $\Phi_{CN,h}$ the Crank-Nicolson map for the remaining terms on a time interval $h$, the time integration is obtained using the second-order Strang splitting:
\begin{equation}
    Q^{n+1} = \left( \Phi_{CN,h/2} \circ \Phi_{iso,h} \circ \Phi_{CN,h/2} \right) Q^n,
\end{equation}
where $h$ is the time step and $n$ the time level.\\ 

In the following, we report the flow generated at different values of Lamb's parameter $\gamma$. In order to appreciate the qualitative changes in the flow, we chose to display three cases: $\gamma = 0$ (which corresponds to the limit $R_d\to\infty$ in which we recover the two-dimensional Navier-Stokes equations), $\gamma = 10^3$ and $\gamma = 10^4$.\\

\begin{figure}
\centering
\includegraphics[width=\textwidth]{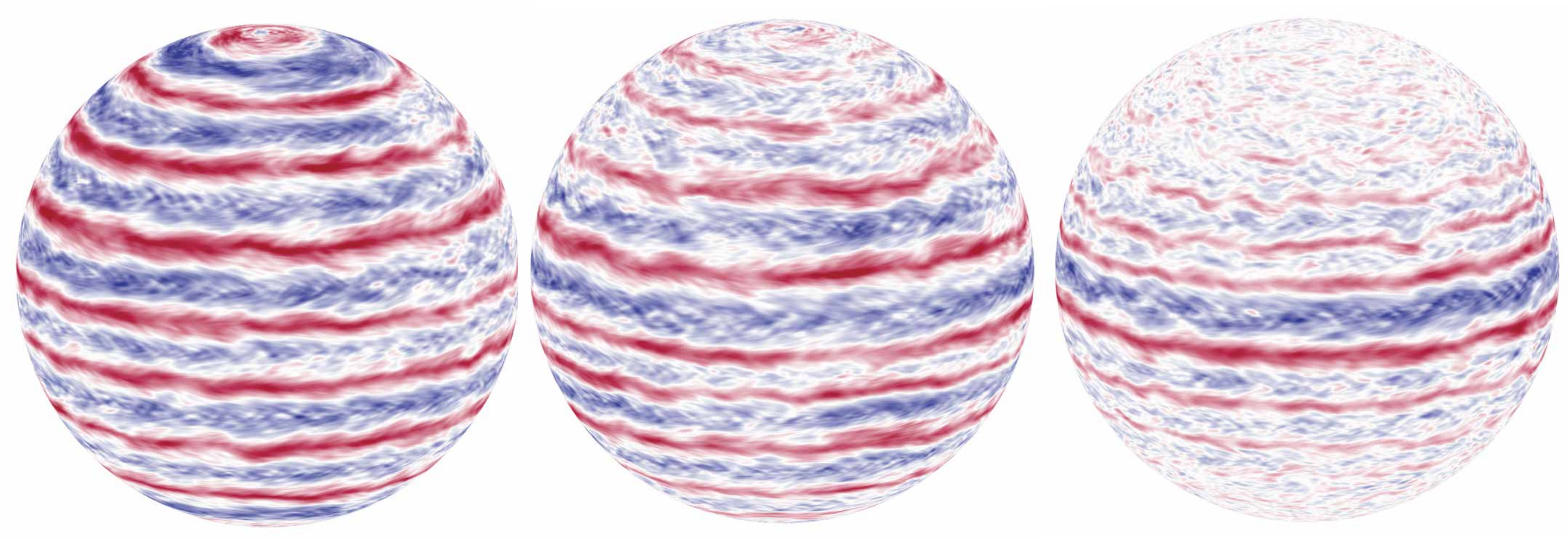}
\caption{Zonal velocity of statistically stationary BSW flow at $\gamma=0$ (left), $\gamma=10^3$ (middle) and $\gamma=10^4$ (right). The colour scale is the same in each simulation and ranges between red and blue, representing \add{prograde} and \add{retrograde} motion respectively}
\label{fig:Spheres}
\end{figure}

Figures~\ref{fig:Spheres} show a snapshot of the zonal component of the velocity field at a time at which the flow is in a statistically stationary state for each choice of parameter $\gamma$. The spatial resolution is set to $N=2048$, which was found to accurately represent the smallest scales in the flow.\\

As a reference, we consider the solution obtained at $\gamma = 0$ as shown in the left snapshot of Figure~\ref{fig:Spheres}. The solution displays a clear zonal structure, with the formation of a large number of zonal jets. In particular, we notice how these jets appear to have an intensity that is quite independent of the latitude. The situation is different when we solve the BSW equation at $\gamma_1=10^3$ (middle snapshot in Fig. \ref{fig:Spheres}). Here, we see that while the number of zonal jets appears unaffected, there is an apparent graded intensity of the jets which ranges from comparably weak jets near the poles to strong jets near the equator. This trend is even stronger when Lamb's parameter is increased to $\gamma = 10^4$ (right image in Fig. \ref{fig:Spheres}), which corresponds to a smaller Rossby deformation length.\\

\begin{figure}

\centering
	\includegraphics[width=.6\textwidth]{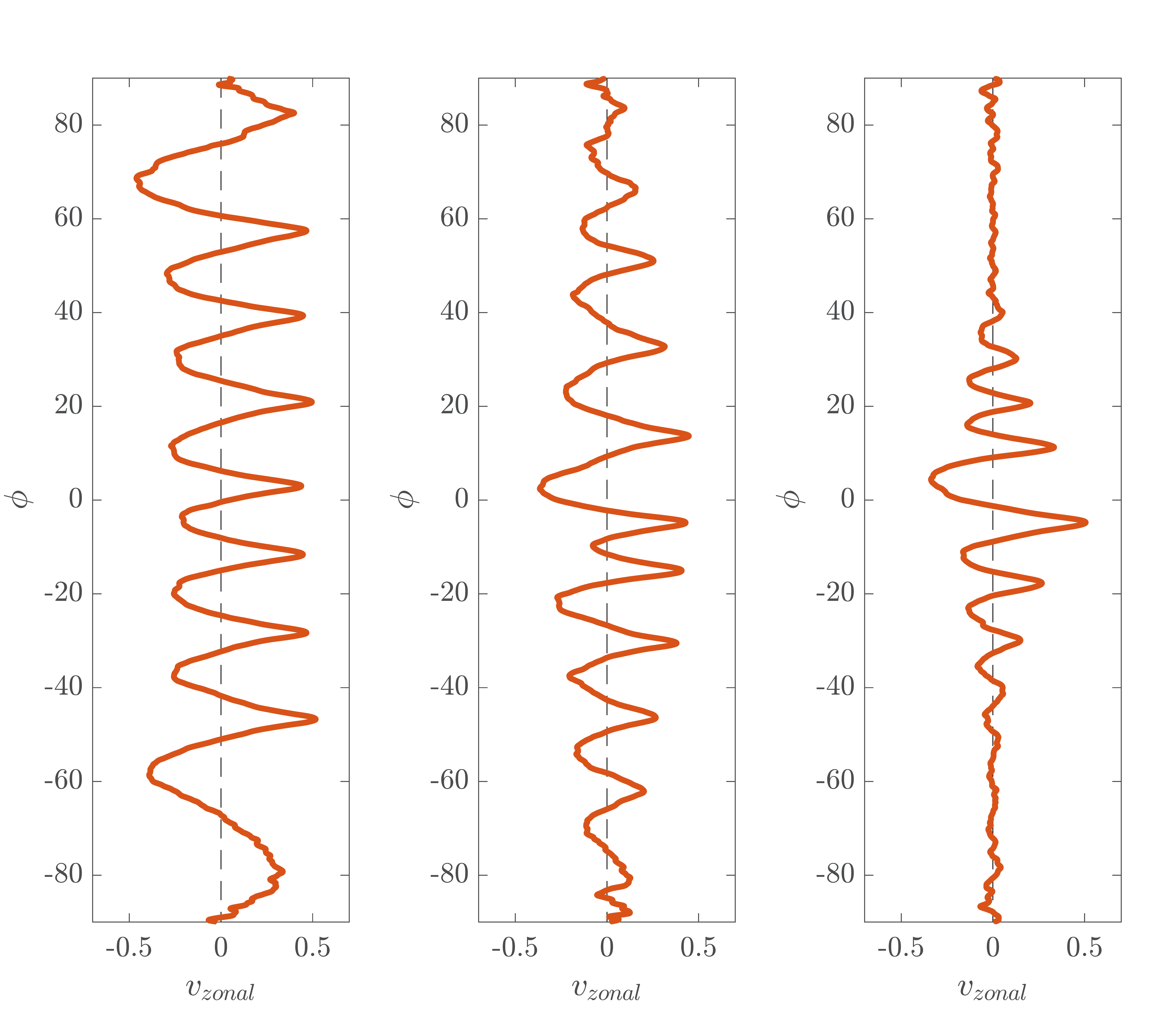}
    \caption{Mean zonal velocity profiles averaged over longitude as a function of latitude for the NS flow (a) and simulations of the BSW equation at $\gamma = 10^3$ (b) and $\gamma = 10^4$ (c).}
    \label{fig:Vzon_profiles}
\end{figure}

This effect is illustrated quantitatively in Figure~\ref{fig:Vzon_profiles}, where the zonal velocity, averaged along longitudes, is displayed as a function of latitude. Here we clearly see how the width and intensity of jets near the equator are almost unaffected by the change in the Lamb parameter. However, we observe that the intensity of jets close to the poles gets attenuated strongly with increasing $\gamma$. \add{This phenomenon of jet attenuation is a physical mechanism which is observed in giant gaseous planets, like Jupiter. The attenuation of zonal jets, while not present in simulations of 2D Navier-Stokes equations (as we see in} Figure~\ref{fig:Vzon_profiles}), \add{is captured by numerical simulations of RSW equations. For example, in the work of }\cite{Scott2007}\add{ it is shown how the equatorial confinement is present in the RSW model and it is regulated by the Rossby deformation radius, similar to the current simulations.}\\

\begin{figure*}
	\includegraphics[width=\textwidth]{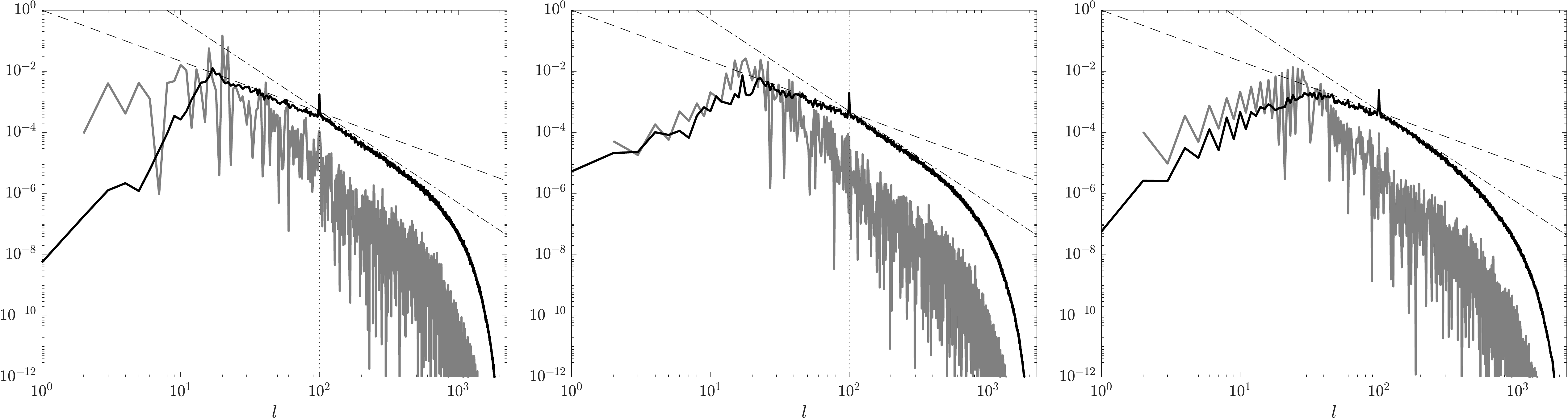}
    \caption{Kinetic energy spectra 
    for the cases $\gamma=0$ (left), $\gamma=10^3$ (middle) and $\gamma=10^4$ (right). The spectra are shown as a function of the spherical harmonics number $l$, split into the zonal modes ($m=0$) in grey, and the non-zonal modes ($m\neq 0$) in black. For reference, the Kraichnan scalings $-5/3$ and $-3$ are shown. The dotted vertical line indicates the modes at which the forcing is centred.}
\label{fig:spectra}
\end{figure*}

In Figure \ref{fig:spectra} we show the kinetic energy spectrum for the three values of Lamb's parameter, in terms of the degree $l$ of the spherical harmonics. The spectra have been split into the zonal ($m=0$) and the non-zonal ($m\neq0$) energy contributions. First of all, we notice how, in all spectra, the non-zonal energy undergoes a double cascade, with a slope of $-5/3$ towards large scales and of $-3$ towards small scales, as expected for 2D turbulence on the sphere \cite{Lindborg2022}. This double cascade was also reported recently for isotropic turbulence in \cite{Cifani2022b}. Also, we notice how the most energetic large-scale modes come from the zonal contribution, with a peak around $l=20$ for all three cases.\\

As we increase the value of $\gamma$, we observe two trends in the kinetic energy spectra which relate to the phenomenon of equatorial confinement of zonal jets. First, we notice that, while the degree $l$ of the peak in the energy spectrum of the zonal modes appears unaffected by $\gamma$, the energy contained in these modes clearly diminishes as $\gamma$ grows. Second, in the case where $\gamma=0$, we notice that there is a rapid decay in the energy of the non-zonal modes as we move towards low wavenumbers $l$. This effect is known as the anisotropic Rhines barrier effect\cite{Rhines1975}, and it is believed to be responsible for the formation of zonal jets \cite{Vallis1993}. This effect takes place at low wavenumbers $l$ with $m \neq 0$, i.e. in the spectral region where the flow is dominated by waves rather than turbulence \cite{Huang1998}.  When $\gamma \neq 0$, the decay of energy in non-zonal modes becomes less rapid, i.e., they contain more energy than their counterparts at $\gamma =0$. As a consequence, even if at low wavenumbers the spectrum is still dominated by zonal modes, the non-zonal modes at low $l$ become important when $\gamma \neq 0$.\\

We conclude by noticing that a similar phenomenon has been studied in the works of Theiss \cite{Theiss2004, Theiss2006}. In the first of these works, a generalized quasi-geostrophic equation that allows a latitude-dependent Rossby deformation length is considered. Similarly to what we find in our simulations of the BSW model, they show an equatorial confinement of zonal jets, together with an attenuation of the Rhines barrier, which can be bypassed at certain latitudes. The similarity between these numerical results can be explained by noticing that in the BSW model, the effect of a variable Rossby deformation length is naturally encoded in the factor $\gamma \mu^2$ in (12).

\section{\label{sec:Conclusions}Conclusions}

\add{In this work, we developed a Casimir preserving numerical method for the BSW equations} (\ref{eq:Lie-Poisson}) \add{on the sphere. Simulations of unforced turbulence in the absence of dissipative terms show that this method approximately conserves energy and conserves Casimirs with a relative error of $10^{-10}$. The BSW model has been simulated in the presence of dissipation and forcing. In this case, we observed that the BSW model is able to capture the phenomenon of jet attenuation, which is shown to be regulated by Lamb's parameter.}\\ 

\add{The presented method for the BSW equations can be readily extended to include a non-trivial bottom topography, since the topography only enters the model as an additional source of potential vorticity in the diagnostic equation }(\ref{eq:potential_vorticity})\add{ for the streamfunction} (see \cite{Verkley2009a})\add{. One application of the present method involves examining the statistical relevance of high-order Casimirs for the BSW model, and their influence on the coarse-grained large-scale flow, following the work of }\cite{Abramov2003}. \add{Furthermore, this modelling approach may be applied to multi-layered models such as the multi-layered quasi-geostrophic equations, which will be the subject of future research.}

\appendix

\section{Scalar function products in the Zeitlin model}
\label{app:basis}
    In this appendix, we add some details regarding the quantized basis $\widehat{T}_{lm}$ and the discrete laplacian $\Delta_N$. \\  
    
    The quantized basis $\widehat{T}_{lm}$ is defined as follows: 
    \begin{equation}
        (\widehat{T}_{lm})_{ij} = (-1)^{\frac{N-1}{2}-i} \sqrt{2l+1}
    \left(\begin{array}{ccc}
\frac{N-1}{2} & l & \frac{N-1}{2} \\
-i & m & j
\end{array}\right)
\label{T}
\end{equation} where the bracket denotes the Wigner 3j-symbols. The normalization is such that $Tr(\widehat{T}_{lm}\widehat{T}^\dagger_{l'm'})= \delta_{ll'}\delta_{mm'}$. Notice that the indices $i,j$ go from $-\frac{N-1}{2}$ to $\frac{N-1}{2}$, having value $(i,j)=(0,0)$ at the "center" of the matrix. \\

The expression for the quantized laplacian has been derived by Hoppe \& Yau \cite{Hoppe1998} and reads:
\begin{equation}
    \Delta_N = \frac{N^2-1}{2} ([\mathbf{X_3^N},[\mathbf{X_3^N},\cdot]] - \frac{1}{2}[\mathbf{X_+^N},[\mathbf{X_-^N},\cdot]] -\frac{1}{2}[\mathbf{X_-^N},[\mathbf{X_+^N},\cdot])
\end{equation}
where $X_3^N \propto \widehat{T}_{10}$ and $X_{\pm}^N \propto \widehat{T}_{1\pm1}$. An important property of $\Delta_N$ is that $\Delta_N\widehat{T}_{lm} = -l(l+1)\widehat{T}_{lm}$ for any $l = 0,...,N$ and $m = -l,...,l$, and its importance is twofold. First of all, it permits the expression of the discrete Laplacian in terms of tridiagonal blocks, as described in Appendix C. Secondly, it permits to compute the basis (\ref{T}) without recurring to the expensive computations of the Wigner 3j-symbols.

\section{The product $\sin^2 \phi \psi$} \label{app:product_projection}
In this appendix, we discuss the approximated law (\ref{eq:approx}). First of all, we notice that if we restrict to the space of polynomials of degree $N$, $Poly_{\mathbb
C}(S^2)_{\leq N } \subset C_{\mathbb
C}^{\infty}(S^2)$, we have that the projection $\Pi_N$ is bijective. The inverse map $\Pi_N^{-1}$ is called "symbol correspondence" and it induces a new product between elements of $Poly_{\mathbb
C}(S^2)_{\leq N }$. This product is called twisted product and reads as follows: if $Y_{lm} = \Pi_N ^{-1}(i\widehat{T}_{lm})$ and $Y_{l'm'} = \Pi_N ^{-1}(i\widehat{T}_{l'm'})$
\begin{equation}
    Y_{lm} \star Y_{l'm'} = \Pi_N ^{-1}(i\widehat{T}_{lm}\cdot \widehat{T}_{l'm'})
\end{equation}
where on the r.h.s. we have the standard matrix product.\\

The important property of the twisted product is that 
\begin{equation}
    \lim_{n \to \infty }(Y_{lm} \star Y_{l'm'}+Y_{l'm'} \star Y_{lm}) = 2Y_{lm}Y_{l'm'}
    \label{twisted}
\end{equation}
where the limit above is taken uniformly, i.e. we have uniform convergence of the sequence of functions on the l.h.s. to the function on the r.h.s \cite{Rios2014}. We remark that this asymptotic ($N \to +\infty$) expansion of the twisted product
is invalid without the assumption $l,l' \ll N$. \\

So we proceed as follows: we approximate the product $\sin^2 \phi \psi$ by means of (\ref{twisted}); then we project through $\Pi_N$. We obtain
\begin{equation}
    \sin^2 \phi \psi \ \ \to \ \ \frac{i}{2}(SP+PS)
\end{equation}
where $P = \Pi_N\psi \in \mathfrak{u}(N)$, $S = \Pi_N \sin^2\phi \in \mathfrak{u}(N)$.\\

Now, we know that the Poisson brackets are approximated by the rescaled commutator $\frac{N^{3/2}}{\sqrt{16 \pi}}[\cdot,\cdot]$ up to an error $O(1/N^2)$. But how well is the product $\sin^2{\phi}\psi$ approximated by $\frac{-i}{2}(PS+SP)$? To compute it we need to use some asymptotics of the 6j Wigner symbol. Taking the inner product of our function product with a basis function, we use the following identity:
\begin{equation}
\begin{aligned}
&\int_0^{2 \pi} \int_0^\pi Y_{l_1}^{m_1}(\theta, \phi) Y_{l_2}^{m_2}(\theta, \phi) Y_{l_3}^{m_3}(\theta, \phi) \sin \phi d \phi d \theta\\
&=\sqrt{\frac{\left(2 l_1+1\right)\left(2 l_2+1\right)\left(2 l_3+1\right)}{4 \pi}}\left(\begin{array}{ccc}
l_1 & l_2 & l_3 \\
0 & 0 & 0
\end{array}\right)\left(\begin{array}{ccc}
l_1 & l_2 & l_3 \\
m_1 & m_2 & m_3
\end{array}\right)
\end{aligned}
\label{eq:sph_trace}
\end{equation}
On the other hand, we have for our renormalized basis \cite{Hoppe1998}
\begin{equation}
\begin{split}
    Tr(\widehat{T}_{l_1m_1}\widehat{T}_{l_2m_2}\widehat{T}_{l_3m_3}) &= \ (-1)^{2s} \sqrt{\left(2 l_1+1\right)\left(2 l_2+1\right)\left(2 l_3+1\right)} \\ &\left(\begin{array}{ccc}
l_1 & l_2 & l_3 \\
m_1 & m_2 & m_3
\end{array}\right)\left\{\begin{array}{ccc}
l_1 & l_2 & l_3 \\
s & s & s
\end{array}\right\}
\label{trace}
\end{split}
\end{equation}
where $s = \frac{N-1}{2}$ and $\{:::\}$ denotes the 6j Wigner symbol. For the latter, there is a known asymptotic formula\cite{Flude1998}: given $a,b,c$ satisfying the triangle inequality and $R \gg a,b,c$ we have
\begin{equation}
\begin{split}
    \left\{\begin{array}{ccc}
a & b & c \\
R & R & R
\end{array}\right\}& =\frac{1}{\sqrt{2 R+1}}(-1)^{(a+b+c)}\left(\begin{array}{ccc}
a & b & c \\
0 & 0 & 0
\end{array}\right) \\&  + O((2R+1)^{-3/2})
\label{eq:asymptotics}
\end{split}
\end{equation}
Which indicates that the right scaling for the approximation is $\frac{i}{2}\sqrt{\frac{N}{4\pi}}(-1)^{2s}(SP+PS)$, with an error of $O(1/N)$. As a final comment, we notice that in case we want to approximate the Poisson Brackets $\{Y_{l_1,m_1},Y_{l_2,m_2}\}$, we have an exact expression in terms of commutators when the $l_1,l_2,l_3$ in (\ref{eq:sph_trace}) have even sum $l_1+l_2+l_3$ (See \cite{Hoppe1998}). When the sum is odd, the first order in the expansion (\ref{eq:asymptotics}) is zero, due to the 3j Wigner symbol property, and for this reason the commutator $[\cdot,\cdot]$ is eventually rescaled by a factor $N^{3/2}$.

\section{Tridiagonal splitting} \label{ap:tridiagonal}
The discrete Laplacian $\Delta_N$ admits a decomposition in tridiagonal blocks, which enables an efficient implementation of the associated eigenvalue problem\cite{Cifani2023a}. In this appendix, we show that the same splitting applies when solving the for the stream matrix $P$ in equation \ref{eq:potential_vorticity}. Since the matrices $\widehat{T}_{lm}$ are eigenvectors of the discrete Laplacian and have entries only in the $m^{th}$ diagonals, we can consider the action of $\Delta_N$ on each diagonal separately. We obtain that $\Delta_N$ decomposes in $(2N-1)$ blocks $\Delta^m$ of size $N-|m|$ with entries, for $m\geq0$,
\begin{equation}
    \begin{split}
        \Delta^m_{ij} &= 2\delta^j_i(s(2i+1+m)-i(i+m)) \\ &-
    \delta^j_{i+1}\sqrt{(i+m+1)(N-1-i-m)} \sqrt{(i + 1)(N -1 -i)} \\ &  - \delta^j_{i-1}\sqrt{(i + m)(N - i - m)}\sqrt{i(N - i)}
    \end{split}
    \end{equation}
where $\delta$ is the Kronecker-delta, $s = (N - 1)/2$ and $i,j = 0,..., N-m-1$. \\

On the other hand, also the term $(PS+SP)$ in (\ref{eq:zeitlin}) can be described off-diagonalwise. Indeed, since the matrix $S$ is diagonal, we find that the $m^{th}$ diagonal of $P$ gets element-wise multiplied by the diagonal matrix  $\mbox{diag}(s_{11}+s_{m+1m+1},s_{22}+s_{m-1m-1}, ..., s_{N-|m|N-|m|}+s_{NN} ) =: \mbox{diag}(\Tilde{S}^m)$.  Thus, we can define a modified discrete laplacian with blocks
\begin{equation}
    \tilde{\Delta}^m_{ij} = \Delta^m_{ij}-4\frac{R^2}{R_d^2} (\Tilde{S}^m)_{ij} \delta_{ij}
\end{equation}
so that equation \ref{eq:zeitlin} can be efficiently solved in the following form:
\begin{equation}
    \begin{cases}
        \Dot{Q} = \left[ P, Q \right]_N \\
        \tilde{\Delta}_NP = Q -2 M
    \end{cases}
\end{equation}

\section*{Open Research Section}

Computations were performed using the open-source Fortran90 package Geometric Lie-Poisson Isospectral Flow Solver (GLIFS) available as \cite{GLIFS_ref}. The simulation data used in this study \cite{SimulationData} can be accessed via https://doi.org/10.5281/zenodo.8116153. All post-processing was done in Matlab \cite{MATLAB:R2019b}. The post-processing software files \cite{PostProcessing} are accessible via \url{https://doi.org/10.5281/zenodo.8112852}.

\paragraph*{Acknowledgments}
\add{The authors would like to thank the associate editor and the anonymous referees for their diligent review and insightful feedback, which greatly contributed to the refinement of this manuscript.} Furthermore, the authors gratefully acknowledge fruitful discussions with Erwin Luesink and Sagy Ephrati (University of Twente), with Darryl Holm (Imperial College London) and Milo Viviani (Scuola Normale Superiore, Pisa). Simulations were made possible through the Muliscale Modeling and Simulation computing grant of the Dutch Science Foundation (NWO) - simulations were performed at the SURFSara facilities in the Multiscale Modeling and Simulation program.

\bibliographystyle{plainnat}
\bibliography{References_final}

\end{document}